\documentclass{aa}
\usepackage{graphicx}
\usepackage[varg]{txfonts}

\def\tex {\ifmmode{{T}_{\rm ex}}\else{$T_{\rm ex}$}\fi}
\def\tmb {\ifmmode{{T}_{\rm mb}}\else{$T_{\rm mb}$}\fi}
\def\ci     {\ifmmode{{\rm C}{\rm \small I}}\else{C\ts {\scriptsize I}}\fi}
\def\hi     {\ifmmode{{\rm H}{\rm \small I}}\else{H\ts {\scriptsize I}}\fi}
\def\hh     {\ifmmode{{\rm H}_2}\else{H$_2$}\fi}

\def\ts     {\thinspace}
\def\kms    {\ifmmode{{\rm \ts km\ts s}^{-1}}\else{\ts km\ts s$^{-1}$}\fi}
\def\msol   {\ifmmode{{\rm M}_{\odot}}\else{M$_{\odot}$}\fi}
\def\lsol   {\ifmmode{{\rm L}_{\odot}}\else{L$_{\odot}$}\fi}
\def\zsol   {\ifmmode{{\rm Z}_{\odot}}\else{Z$_{\odot}$}\fi}

\begin{document}

\title{Discovery of CO absorption at z=0.05 in G0248+430
\thanks{Based on observations carried out with NOEMA, the NOrthern Extended Millimeter Array -- IRAM (Institute of RAdioastronomy in Millimeter).  } 
}

\author{F. Combes \inst{1}
\and
N. Gupta \inst{2}
\and
G. I. G. Jozsa \inst{3,4,5}
\and
E. Momjian \inst{6}
          }
\institute{Observatoire de Paris, LERMA, College de France,
CNRS, PSL Univ., Sorbonne University, UPMC, Paris, France
 \and  
Inter-University Centre for Astronomy and Astrophysics, Post Bag 4, Ganeshkhind, Pune 411 007, India
 \and 
South African Radio Astronomy Observatory, Black River Park, 2 Fir Street, Observatory, Cape Town, 7925, South Africa
 \and 
Rhodes Centre for Radio Astronomy Techniques \& Technologies, Dep. of Physics and Electronics, Rhodes University, PO Box 94, Grahamstown 6140, South Africa
 \and 
Argelander-Institut f\"ur Astronomie, Auf dem H\"ugel 71, D-53121 Bonn, Germany
\and 
National Radio Astronomy Observatory, P.O. Box O, Socorro, NM 87801, USA
        }

   \date{Received  2018/ Accepted  2019}

   \titlerunning{CO absorption at z=0.05}
   \authorrunning{F. Combes et al.}

     \abstract{ Absorption lines in front of distant quasars are quite rare
     in the millimeter domain. They can, however, bring  very useful and
     complementary information to emission lines.
     We report here the detection with NOEMA of CO(1-0)  and CN(1-0) lines
     in absorption, and the confirmation of CO emission in the quasar/galaxy
     pair Q0248+430/G0248+430. The system G0248+430 corresponds to two merging galaxies
     (a Seyfert and a LINER) at z=0.0519 with a tidal tail just on the line of sight
     to the background quasar Q0248+430 at z = 1.313.
     Optical (CaII, NaI), \hi-21cm, and OH-1667 MHz absorption lines 
     associated with the tidal tail of the foreground system have 
     previously been detected toward the quasar, while four CO lines
     at different rotation $J$ levels have been detected in emission from
     the foreground galaxies.
     New \hi\, 21cm line observations with the upgraded GMRT array are also presented.
     We discuss the molecular content of the merging galaxies, and the
     physical conditions in the absorbing interstellar medium of the tidal tail.
}

\maketitle


\section{Introduction} 
\label{sec:intro}  

Most of our knowledge of molecular gas in galaxies at high or low redshift have been obtained through CO emission
line studies. However, absorption lines can bring new and complementary information. In contrast to emission,
absorption lines remain observable at practically any distance, with the sensitivity only determined by the
strength of the background source  \citep[e.g.,][]{Combes2008}.
Given a strong enough continuum source, ($\gtrsim 50$~mJy), millimeter-wave absorption lines can be used to
obtain information about molecular gas and star formation in ``normal'' galaxies. While emission lines are
sensitive to dense and warm molecular gas, absorption lines may also arise from low-excitation and diffuse gas,
which is more prevalent in normal galaxies 
 \citep[e.g.,][]{Wiklind1995, Wiklind1996, Wiklind1997, Menten2008, Henkel2009,
 Muller2014}. 

 Once a galaxy has been detected in the strongest CO absorption lines, deeper studies in other
 molecular lines allow characterization of  the physical and chemical conditions in the absorbing gas 
\citep[e.g.,][]{Henkel2005, Bottinelli2009, Muller2014, Muller2016, 
Riquelme2018}.
The relative strengths of species like HC$_3$N, where the excitation is dominated by the cosmic
microwave background (CMB), can be used to determine the CMB temperature 
\citep[e.g.,][]{Henkel2009, Muller2013}.
Comparisons between the redshifts of different transitions (e.g., NH$_3$,
CH$_3$OH, OH) can be used to test for cosmological variations in the fundamental constants 
\citep[e.g.,][]{Uzan2011, Kanekar2011, Kanekar2012, Rahmani2012, Bagdonaite2013}.\\

In this paper we report the discovery of CO absorption at $z\sim$0.05,
 in front of the quasar Q0248 at $z=1.31$ \citep{Kuehr1977}. 
 The absorption comes from a foreground pair of galaxies called G0248
 \citep{Junkkarinen1987}, consisting of the two sources G1 and G2. 
   A tidal tail connects, in projection, G2 with the background quasar. Optical absorption lines
   (CaII and NaI) have been detected at the G0248 redshift $z=0.05194$, or very close  (0.05146, at
   $\sim$150~\kms) by 
 \cite{Sargent1990} and \cite{Womble1990}. They noted that the QSO and the galaxy nuclei are
 separated by 14.7'' = 14.7~kpc at z=0.05, and the absorption is due to the tidal tail crossing the
 quasar (see  Fig. \ref{fig:ps1}). 

Throughout this paper we use the $\Lambda$CDM cosmology with $\Omega_m$=0.3, $\Omega_\Lambda$=0.7, and 
H$_{\rm o}$=70\,\kms\,Mpc$^{-1}$. At the redshift of the foreground galaxies, the distance scale is 1~kpc per arcsec.

\section{Apparent quasar-galaxy merger association}
\label{sec:q0248}   
 The associated pair of galaxies G0248 is a violent starburst, 
 according to its IRAS far-IR (FIR) flux density measurements, corresponding to 
 L$_{\rm FIR}$= 4.3$\times$10$^{11}$ L$_\odot$, and a star formation rate  (SFR) of 74 M$_\odot$/yr (see Table \ref{tab:G0248}), 
 and the  G1 and G2 nuclei (separated by 3.5’’ = 3.5~kpc) both show nonthermal 
 activity \citep{Borgeest1991, Kollatschny1991}. Indeed, G1 is a LINER
 and G2 a Seyfert 2, and based on their optical emission line excitation
 both are located in the AGN region of the 
  Baldwin-Phillips-Terlevich (BPT) diagram
  \citep{Baldwin1981}, plotting [OIII]/H$\beta$ as a function of [NII]/H$\alpha$.
 We note that the above FIR luminosity is
 extrapolated between 1 and 500$\mu$m, as in Table \ref{tab:G0248},
 while \cite{Gupta2018OH} adopt the {calibration formula} from \cite{Sanders1996}
 and find L$_{\rm FIR}$= 3.5$\times$10$^{11}$ L$_\odot$, 
 which is in good agreement.
 The background quasar Q0248 is a well-studied radio source and has a  variable, flat spectrum 
 \citep{Massaro2014}.
 Its flux density at 100 GHz has been observed to be around 200~mJy.
  The variability of the quasar could be due in part to micro-lensing by stars 
 in the tidal tail of G0248 \citep{Borgeest1994}. 
 An OH megamaser has been detected from the starburst 
 \citep{Kazes1989, Darling2002}. \cite{Kollatschny1991}
 determine the kinematics of the two merging galaxies,  the LINER G1 and Seyfert G2; 
 the former shows systematic rotation, but  the latter does not.  With respect to G2 (z=0.0507),
 a constant velocity of $\sim$ 100 \kms\ is observed all across the tidal tail.
 Furthermore, from their spectra,  they conclude that G1 is a spiral galaxy and G2 an 
 elliptical. \cite{Downes1993} report the detection of CO(1-0) emission, 
 with an integrated flux of 25 Jy \kms\ with the IRAM 30m telescope. 
   With the NRAO 12m telescope, \cite{Walker1997}
 confirm this detection, and show a quite high CO(2-1) 
 excitation, with a ratio of $\sim$ 4 between the CO(2-1) and CO(1-0) 
 integrated fluxes, taking into account a common full width at half maximum (FWHM) of 175~\kms\,. 
 Observations of CO(3-2) with the JCMT (James Clerk Maxwell Telescope) by \cite{Leech2010} reveal a 
 CO(3-2)/CO(1-0) flux ratio of $\sim$ 6. \cite{Papadopoulos2010} find a 
 CO(6-5)/CO(1-0) flux ratio of 16, showing that the peak of the excitation 
 is around J=3-4. \cite{Hwang2004} have mapped the galaxy pair in CO(1-0) 
 at 2’’ resolution using the Berkeley Illinois Maryland Array (BIMA). 
 They confirm the integrated flux of 25 Jy \kms\ 
 found by \cite{Downes1993} showing little or no  interferometer filtering 
 (see our spectrum in Fig. \ref{fig:specabs}). They also show \hi-21cm 
 absorption detected with the Very Large Array (VLA, 1.8'' beam) 
 in front of the quasar, which 
 remains unresolved in the 21cm continuum. The \hi\, absorption is at 
 the same velocity as the CaII 
 and NaI lines. \cite{Gupta2006, Gupta2018OH} have detected the strongest 
 line (at 1667 MHz rest frequency)
 of the OH 18cm transitions in absorption in front of the quasar with a total integrated
 optical depth of 0.08$\pm$0.01~\kms\,, showing that molecular gas exists in 
 the tidal tail.
For an excitation temperature of 10 K, the observed 1667 MHz OH optical depth 
corresponds to a column density of N(OH) = 1.8 x 10$^{14}$ cm$^{-2}$;
similar column densities (10$^{13 -14}$ cm$^{-2}$) are observed in diffuse clouds,
while typical values for giant molecular clouds (GMC) are 10$^{15 - 16}$ cm$^{-2}$.
 The BIMA observations by \cite{Hwang2004} did not lead to any detected CO absorption in 
front of the quasar, but the data suffered from a low sensitivity 
(N(CO) $< 10^{16}$ cm$^{-2}$).

 Assuming a spin temperature of $\sim$ 1000K, \cite{Hwang2004} derive a 
 column density of N(\hi) = 5$\times$10$^{20}$ cm$^{-2}$.
 However, this is based  on a derived integrated optical depth at 21cm of
 $\int \tau_{21}$ dv = 0.26~\kms. \cite{Gupta2018OH} reprocessed these archival
 VLA data, and found $\int \tau_{21}$ dv = 0.43$\pm$0.02~\kms.
 The column density of the \hi\, gas in front of the quasar is large 
 enough for the absorber to be classified as a damped Ly$\alpha$ system,
provided the spin temperature is higher than 300 K and/or the covering factor is less than unity.

 The calcium depletion on grains of the absorbing gas is high, indicating that
the gas must come from a disk rather than from a halo.
 The column density of NaI is 6.3 $\times$ 10$^{13}$ cm$^{-2}$ \citep{Womble1990}, while 
 the ratio N(CaII)/N(NaI)=0.2-0.3 is low, similar to or lower than the values $\sim$ 1 of the Galactic disk. 
 The high NaI suggests that the gas could come from outflowing gas from the 
 starburst or the active nuclei in the tidal tail \citep{Heckman2000}. The absence of \hi\, emission
 in the VLA observations (upper limit at 5$\sigma$ of 
 2.3$\times$ 10$^{22}$ cm$^{-2}$ over 30~\kms) is likely due to the phase 
 transformation of the atomic gas to molecular gas in the starburst, which 
 now contains M(H$_2$)= 1.6$\times$10$^{10}$ M$_\odot$, with a standard 
 CO-to-H$_2$ conversion factor, X(CO) = 2$\times 10^{20}$ cm$^{-2}$ / (K \kms).

\section{Observations and data analysis}      
\label{sec:obs}   

\subsection{IRAM observations}      
\label{sec:iramobs}   

We have mapped G0248 at 109.605~GHz for a total of 13~hr (8~hr on source) allowing us
to spatially resolve the CO(1-0) emission line in the foreground gas,
and the continuum emissions from both the foreground merging system 
and the background quasar.

The phase center was RA(2000)~=~02h 51m 35.1s, Dec(2000)~=~43$^\circ$ 15' 14.0'', corresponding to
the barycenter of the quasar and the merging system (both are located at $\sim$ 6.5'' from the phase center).
The observations were made with nine antennas using the extended array
A-configuration of the NOEMA interferometer (project W17CE, P.I.:~Combes). The configuration provided
a synthesized beam of 0.94 $\times$ 0.67 arcsec (PA=60$^\circ$). This allows us to
resolve the two merging galaxies, which are separated by 3.5''= 3.5~kpc.
The field of view at half power is 47'' at this frequency.
The observations were carried out during five days (10, 16, 23, 25~February, and 2 March 2018) in very good 
weather conditions (seeing in the range $\sim$0.1-0.4~arcsec).

We employed the PolyFix correlator in Band 1 (3~mm), which provides 2$\times$8~GHz of instantaneous
dual-polarization bandwidth. The spectral resolution was 2 MHz (5.4~\kms\ at the redshifted CO(1-0) frequency).
The bandwidth allowed us to observe the frequency range
from 91.9 to 99.7 GHz in LSB (Lower SideBand), and 107.3 to 115.1 GHz in the USB 
(Upper SideBand).
The data reduction was performed using the latest release of the GILDAS package as of
May~2018\footnote{https://www.iram.fr/IRAMFR/GILDAS/}.
The data were calibrated using the NOEMA standard pipeline
adopting a natural weighting scheme to optimize both sensitivity and resolution.
Flagging was required only on the February 10 data, and was minor.
The rms noise is 0.3~mJy in 30~\kms\ channels
for the line and 8~$\mu$Jy/beam for the continuum.

\subsection{uGMRT observations}      
\label{sec:gmrtobs}   

We used {\tt Band-5} (1000-1450\,MHz) of the upgraded Giant Metrewave Radio Telescope (uGMRT) to observe the 
redshifted \hi\, 21cm line toward the quasar. The observations took place on 
29 June 2018. 
The GMRT Software Backend (GSB) was used to configure a baseband bandwidth of 4.17\,MHz split
into 512 spectral channels 
(resolution$\sim$1.8\,\kms) centered at 1350.8\,MHz. 
During the $7.5$-hour  observing run,  3C48 was observed for flux density 
scale and bandpass calibrations.  
The total on-source time was 6.6 hours.  The data were edited, calibrated, and imaged using the
Automated Radio Telescope Imaging Pipeline 
({\tt ARTIP}) that is being developed to perform the end-to-end processing of data from the uGMRT
and MeerKAT absorption line 
surveys \citep[][]{Gupta2018HI, Sharma2018}.

\begin{figure} 
\centerline{\vbox{
\centerline{\hbox{ 
    \includegraphics[angle=0,width=8.5cm]{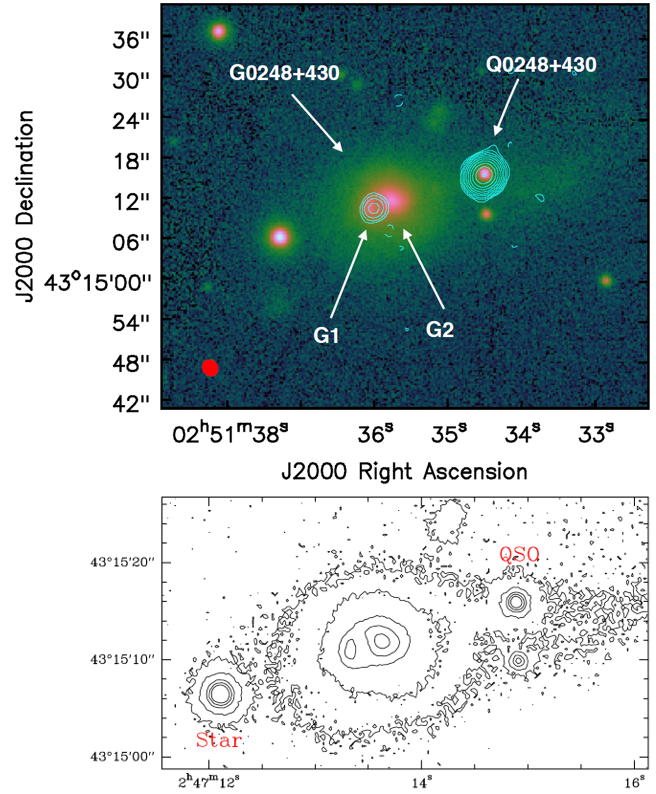}
}} 
}}  
\vskip+0.2cm  
\caption{ {\bf Top}:
   uGMRT radio continuum (1.35\,GHz) contours overlaid on the PS1 $r$-band image.
  The contour levels are 2.2$\times$2$^n$\,mJy\,beam$^{-1}$ 
  (where n=-1,0,1,2,3,...).  The restoring beam of 2.49$^{\prime\prime}\times$2.22$^{\prime\prime}$
  with position angle -35.6$^\circ$ is shown in the bottom 
  left corner. The strong source at the SE of G0248+430 is a foreground star.
  {\bf Bottom: The PS1 $r$-band image in contours, to better show the tidal tail, in
        front of the quasar.}
} 
\label{fig:ps1}   
\end{figure} 

\begin{figure} 
\centerline{
\includegraphics[angle=0,width=8.0cm]{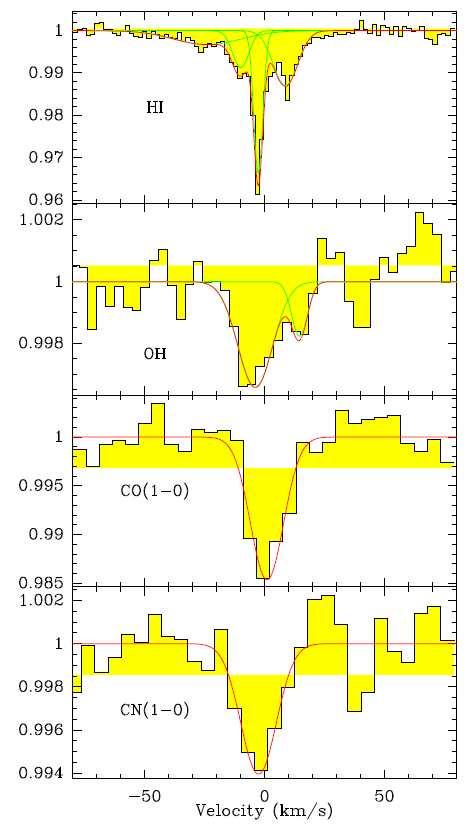}
}
\caption{ \hi\ 21cm, OH (1667 MHz), CO(1-0), and CN(1-0) absorption spectra
  toward the quasar Q0248+430. 
The zero of the velocity scale ($z=0.05151$) is centered at the peak of the low resolution
        \hi\ 21cm absorption line \citep{Gupta2018OH}.
The vertical scale is flux normalized to the continuum. 
} 
\label{fig:specabs}   
\end{figure}

The Stokes $I$ radio continuum image made using the line-free frequency 
channels and using {\tt ROBUST=0.5} visibility weighting (based on Common Astronomy Software Applications, CASA)
 is shown in 
Fig.~\ref{fig:ps1}. The image has a synthesized beam of $2.49^{\prime\prime}\times2.22^{\prime\prime}$ and
an rms noise level of 0.45\,mJy\,beam$^{-1}$. 
The radio continuum emission associated with Q0248+430 and G1 is unresolved with a deconvolved
size $<$0.2$^{\prime\prime}$. The integrated 
flux densities are 1214$\pm$1.3 and 23.7$\pm$1.3\,mJy, respectively.
We extracted the Stokes $I$ 21cm absorption spectra toward Q0248+430 and G1.   
The spectral rms noise level in the unsmoothed spectra is 1.2\,mJy\,beam$^{-1}$.
The spectrum of Q0248+403 is presented in Fig.~\ref{fig:specabs}.   Ninety percent of the total 21cm optical
depth is contained within 49\,\kms\ and 
the total integrated 21cm optical depth, $\int\tau dv$ = 0.53$\pm$0.02\,\kms. In the spectrum of G1
(not shown here), we detect a broad  absorption feature 
$\sim$100\,\kms\  in width centered at 1351.06\,MHz.
It corresponds to an integrated optical depth of
19$\pm$1\,\kms. However, in the different circular polarizations
  (L for left, R for right) and their cross-correlations, this feature 
is present only in LL and not in RR; therefore, we consider
the feature to be an artifact. For G0248+430 we then use
the RR spectrum smoothed to 30\kms.
Adopting 100\,\kms\ as the FWHM of a typical associated 21cm absorption line \citep{Gupta2006}, we estimate a
3$\sigma$ 21cm integrated optical depth limit of 6\,\kms.

\begin{table}[h]
\caption[]{Properties of the merging system G0248}
\label{tab:G0248}
\centering
\begin{tabular}{ccccc}
\hline\hline
$L_{\rm FIR}$ & $M_\star$  &  SFR & $L'_{CO}$ & $M_{H2}$   \\
(L$_\odot$) & { (M$_\odot$)} &  { ($M_\odot/{\rm yr}$)} & 
 {\small (K~km~s$^{-1}$~pc$^2$)} & (M$_\odot$)  \\
 (1) & (2) & (3) & (4) & (5)   \\
 \hline
 4.3$\times10^{11}$ & 6.5$\times10^{10}$ & 74 &  0.4$\times10^{10}$  &  1.6$\times10^{10}$   \\
  & & & &  \\
\end{tabular}
\begin{list}{}{}
\item (1) total infrared luminosity derived from IRAS fluxes \citep{Kollatschny1991}.\\
        (2) stellar mass estimated from the L$_B$ luminosity \citep{Bell2001}.\\
(3) star formation rate estimated using the FIR luminosity, with the relation
SFR = L$_{FIR}$/(5.8$\times$10$^9$ L$_\odot$) compiled by \cite{Kennicutt1998}.\\
(4) Integrated CO line luminosity estimated from the observed CO(1-0) flux \citep{Downes1993}.\\
(5) Total molecular gas mass estimated assuming a CO-to-H2 conversion factor
$\alpha_{\rm CO}=4.36~M_\odot~({\rm K}~{\rm km}~{\rm s}^{-1}~{\rm pc}^{2})^{-1})$.
\end{list}
\end{table}

\begin{table}[h]
\caption[]{NOEMA observations of absorption and emission toward Q0248}
\label{tab:absem}
\centering
\begin{tabular}{lcc}
\hline\hline
          & Absorption  & Emission   \\
RA        & 02:51:34.5       & 02:51:36.04   \\
Dec       & 43:15:16.0       & 43:15:10.8    \\
z         & 0.05151          &  0.05135      \\
 \hline
 S$_{CO(1-0)}$ & 0.25$\pm$.02     &  29$\pm$0.5    \\
 $\Delta$V$_{CO}$ & 16$\pm$1.6     &    191$\pm$2     \\
 S$_{CN(1-0)}$ & 0.11$\pm$.02     &                  --        \\
  $\Delta$V$_{CN}$ & 17$\pm$3     &        --                     \\
\end{tabular}
\begin{list}{}{}
\item -- The absorption position is that of the quasar, the emission position is that of the CO(1-0) peak
  close to G1 (see Fig. \ref{fig:ps1}).\\
-- The integrated absorption signal is the integrated optical depth in \kms\,.
In emission, it is the integrated CO flux in Jy \kms\ for the whole galaxy\\
-- The $\Delta$V are FWHM, in \kms\ \\
\end{list}
\end{table}

\section{Results}    
\label{sec:res}  

\subsection{Implication of CO absorption}    
\label{sec:COabs}

The CO(1-0) emission and possible absorption was mapped in the NOEMA field of view
of 47 arsec FWHP around the merging galaxies (G1 and G2 of Fig 
\ref{fig:ps1}) with 0.94''$\times$0.67'' resolution. Although the quasar
and the tidal tail were only at an angular distance of 6.5'' from the phase center,
the sensitivity was not sufficient to detect molecular emission in the tidal tail.

While it was quite easy to detect the radio continuum of the quasar, at a level of 0.155 Jy 
(and the foreground absorption toward it) it was not possible to detect
 the two AGN nuclei in the continuum
of  G2 (Seyfert) or of GI (LINER), the latter being detected at centimeter wavelength. 

The quasar continuum source is unresolved. In the millimeter range, AGN continuum
sources are in general restricted to a core  smaller than a milliarcsec in size, contrary to
the cm emission, extended due to steep spectrum radio lobes \citep[e.g.,][]{deZotti2010}. This size
corresponds to 1~pc at the galaxies'
distance, and it is therefore justified to assume that the molecular medium fills the surface of
the quasar continuum source.
Assuming a filling factor of 1, we can derive the average column density
over the beam corresponding to the quasar mm continuum emission
and its footprint on the galaxy tidal tail. This will be a lower limit
to the actual column density,

\begin{equation}
 N_{tot} = \frac{8\pi}{c^3} \frac{\nu^3}{g_J A_{J,J+1}} f(T_x) \int{\tau dv}, 
\end{equation}

\noindent where $g_J$ is the statistical weight of level $J$, $A_{J,J+1}$ is the Einstein coefficient for
transition $J \rightarrow J + 1$, and the function $f (T_x )$ is
\begin{equation}
f (T_x ) = \frac{Q(T_x ) e^{E_J /kT_x}}{1 - e^{-h\nu/kT_x}}
\end{equation}

We  adopt the partition function of local thermal equilibrium (LTE),
i.e.,  $Q(T_x ) = \Sigma g_J e^{-E_J /kT_x}$,
where $E_J$ is the energy of level $J$ or $N=J+F$ (for CN) and $T_x$ is the
excitation temperature of the CO or CN molecule.
 For the CN molecule, we detect the two stronger lines at rest frequencies,
  113.488 GHz and 113.490 GHz, which are blended at our spectral resolution
  (they correspond to J=3/2-1/2, F=3/2-1/2, and F=5/2-3/2). We do not detect
  the three other lines, expected to be five times weaker.
We have detected CO(1-0) absorption in front of the quasar,
 with an optical depth $\tau$=0.016 or $\tau \Delta$V =0.25~\kms\,,
 corresponding to N(CO) = 2.9$\times$10$^{15}$ cm$^{-2}$ or
 N(H$_2$) = 2.9$\times$10$^{19}$ cm$^{-2}$, for a common
 CO/\hh\ abundance ratio of 10$^{-4}$. For this computation,
 we have assumed an excitation temperature of $T_x$ = 15~K.
 With the same assumptions, the column density of CN is 7.6$\times$10$^{12}$ cm$^{-2}$.
 The relative abundance of CN/CO = 2.6 10$^{-3}$ is relatively high for dense clouds, by
 1 or 2 orders of magnitude \citep[e.g.,][]{Leung1984}, but not necessarily for 
 more diffuse clouds, for which some models predict abundances
 100 times higher \citep{Wakelam2015}.  The H$_2$ column density found toward
 the tidal tail indeed corresponds  to that of a diffuse interstellar medium \citep{Welty2006}
 (see also Sect. \ref{sec:intro}).

\begin{figure} 
\centerline{
\includegraphics[angle=0,width=8.5cm]{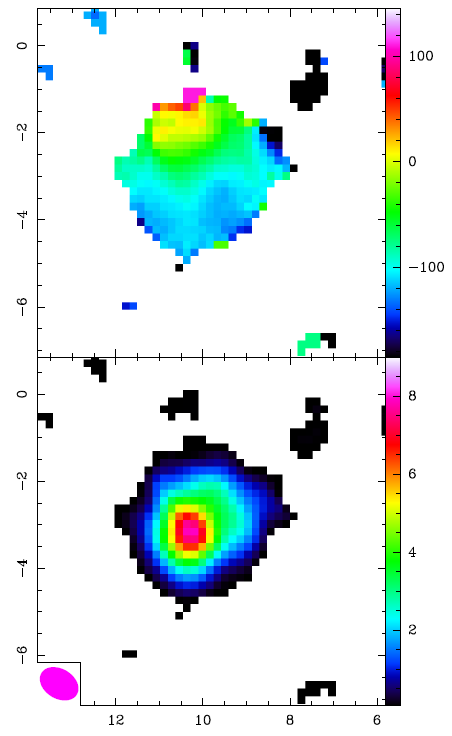}
}
\caption{
CO(1-0) integrated intensity map of the galaxy G0248 (bottom) 
        seen in emission and 
its velocity field (top) from our NOEMA observations.
The scales are in arcsec, with respect to the phase center
RA(2000)= 02h 51m 35.1s, Dec(2000)=43$^\circ$ 15' 14.0'',
and the box size is 8''.
The beam is indicated in the bottom left corner.  
        The integrated flux is in Jy.beam$^{-1}$ \kms\,, and the velocity scale
 in \kms\,, relative to z=0.05151.
} 
\label{fig:CO10map}   
\end{figure} 

\subsection{Implications of CO emission}    
\label{sec:COem}

\begin{figure} 
\centerline{
\includegraphics[angle=0,width=7.0cm]{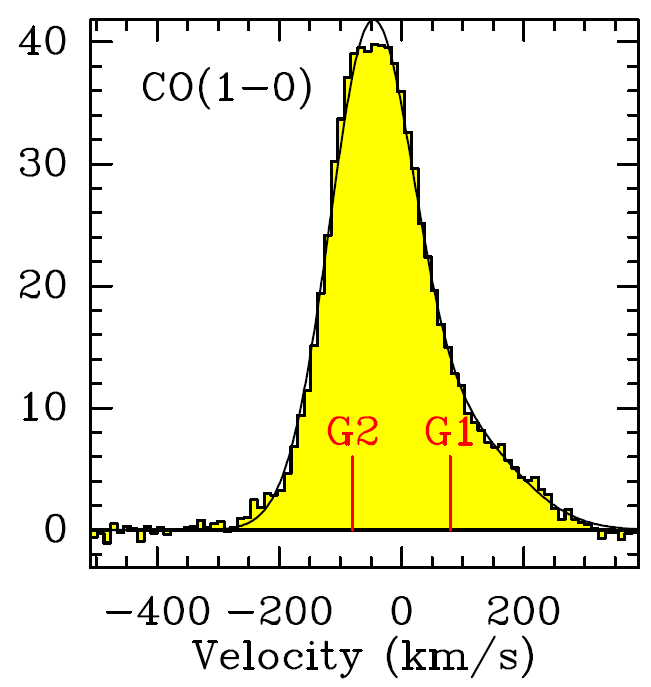}
}
\caption{CO(1-0) emission spectrum toward the GO248 galaxy center. 
The zero of the velocity scale ($z=0.05151$) is centered at the peak of the low
resolution \hi\ 21cm absorption line \citep{Gupta2018OH}.
The systemic velocity of the two merging galaxies G1 and G2 are indicated.
The vertical scale is in mJy\,beam$^{-1}$.
} 
\label{fig:specem}   
\end{figure}

Figure \ref{fig:CO10map} displays the NOEMA integrated CO(1-0) emission map
and the velocity field of G0248 (zero and first moments of the cube).
The central CO(1-0) spectrum is displayed in Fig. \ref{fig:specem} .
For the first time, the CO emission is resolved by our beam of
0.94$\times$0.67'' and the total extent of the CO emission is $\sim$ 4~kpc.
The morphology is  not symmetric, however, as expected for
an interacting galaxy. The velocity field shows clear rotation,
with the kinematic major axis in the N-S direction.
The projected gradient is relatively low, but this can be
explained by the almost face-on orientation (inclination $\leq 30^\circ$)
of both  G1 and G2.
The systemic velocities  of the two galaxies, obtained through
their optical spectrum by \cite{Kollatschny1991}, are indicated in
figure \ref{fig:specem}. They are symmetrically located at $\pm$ 80~ \kms\,
from the center of the absorption, which is at z=0.05151.

With our high spatial resolution, we can now attribute the CO emission
to one or the other galaxy of the pair. The galaxies are separated by 3.5'' = 3.5~kpc,
and  \cite{Hwang2004} find the CO emission centered in the middle of G1 and G2,
but they had a synthesized beam of 2''. As displayed in Table \ref{tab:absem},
the center of the CO emission coincides clearly with the position of the spiral galaxy G1, and
not with the elliptical G2, or with a position in between the two merging
galaxies. Within its maximum radial extent
of 2~kpc, there is no CO emission detected toward the center of G2.
\cite{Kollatschny1991} estimated the V luminosities of both galaxies,
L$_V$(G1) = 4.5$\times$10$^9$ L$_\odot$ and L$_V$(G2) = 9.4$\times$10$^9$ L$_\odot$.
Through kinematical arguments, they also derived a total mass for the
ensemble of M$_{tot}$ = 6.7$\times$10$^{10}$ M$_\odot$. Assuming comparable
mass-to-luminosity ratios for the two galaxies, we can estimate the stellar mass
of G1 at 2.2$\times$10$^{10}$ M$_\odot$.  This estimation is  compatible
with that obtained from the observed H$\alpha$ and [NII] rotation curve: a maximum rotational
velocity of 100~\kms\ at a radius of 2~kpc, provided that the inclination
of the spiral galaxy G1 is 27$^\circ$.

The observation of the fundamental CO(1-0)
line is the best measure of the total \hh\, mass.
We compute L'$_{CO}$, the CO luminosity in  units of K \kms pc$^2$,
through integrating the CO intensity over the velocity profile, and over the galaxy
extent. The total flux is $S_{CO}dV$= 29 Jy km.s$^{-1}$ (see  Table \ref{tab:absem})
very close to the previous values, 24 Jy \kms\ with the BIMA interferometer \citep{Hwang2004}
and  25 Jy \kms\ with the IRAM 30m telescope \citep{Downes1993}, confirming
that the interferometer has not resolved out any emission.

This CO luminosity is given by 
\begin{equation}
L'_{CO} =
3.25 10^7 S_{CO}dV  {{D_L^2}\over {\nu_{rest}^2(1+z)}} \hskip6pt \rm{K\hskip3pt
  km \hskip3pt s^{-1}\hskip3pt pc^2},
\end{equation}
where  $S_{CO}dV$ is the integrated flux  in Jy \kms\,, $\nu_{rest}$ the rest
frequency in GHz,  and $D_L$ the luminosity distance in Mpc.  
Under the  assumption of a standard CO-to-H$_2$ conversion factor \citep{Bolatto2013},
we compute the H$_2$ mass using M$_{\rm H_2} = \alpha$ L'$_{\rm CO}$,
with $\alpha=4.36$ M$_\odot$ (K \kms\, pc$^2$)$^{-1}$.
The molecular gas mass is then M$_{\rm H_2}$ = 1.5$\times$10$^{10}$ M$_\odot$.

Given our estimation of the stellar mass of the spiral galaxy G1 above,
we can now compute a gas fraction of 40\%. This is a high gas fraction,
even for the $z=0.05$ epoch, as determined from the scaling relation
of the main sequence \citep[e.g.,][]{Tacconi2018}. The gas fraction in
G1 is ten times higher than the value expected  on the main sequence.
The explanation is that  G1 belongs to a merging system. The SFR is 74 M$_\odot$/yr (see Table \ref{tab:G0248}), and the depletion time
is t$_{dep}$ = 200 Myr, i.e.,  ten times lower than the depletion time on the main sequence.
The merging system is therefore in a starburst phase, probably due to the
gravitational torques of the interaction, which have driven  all the gas reservoir
of the G1 spiral galaxy inward.
 Alternatively, as observed in active starburst galaxies,
it is possible to adopt a much lower CO-to-H$_2$ conversion factor,
but then the depletion time would be even lower, which would bring the system
even farther from the main sequence.

\subsection{\hi\ absorption}    
\label{sec:HIabs}  

The updated \hi\ 21cm absorption spectrum toward the quasar is presented in Fig. \ref{fig:specabs},
together with the other absorption lines. We have fitted the spectrum with four Gaussians,
and the results are in Table \ref{tab:absHI}. The new \hi\ absorption spectrum
with high spectral resolution allows the discovery of  quite narrow components (8~\kms\,,
implying a kinetic temperature lower than 1400 K),
together with relatively wide wings (33~\kms), although the fit is not unique.
Within the errors, the integrated 21cm optical depth estimated using this spectrum is
consistent with that from the VLA spectrum presented in \cite{Gupta2018HI}.  As expected the
molecular absorption is coincident with the narrower (implying colder) and stronger 21cm absorption
components at -2.6 and +8.7 \kms\,.
The limited spectral resolution and sensitivity of the molecular data 
(OH, CO, and CN) does not allow us to carry out a more detailed
decomposition. The two main and broader components seen in the \hi\ absorption
are only tentatively seen in the OH
spectrum, and not in CO or CN. This might be explained by the different sizes of the quasar
continuum emission at millimeter and centimeter wavelengths.
The radio continuum at 2.3 GHz has an overall extent of 26 mas (27 pc at z=0.05;
\cite{Fey2000}), which is much larger than the size expected at $\sim$ 100 GHz relevant for
CO and CN absorption lines.

\begin{table}[h!]
\caption[]{uGMRT observations of absorption lines}
\label{tab:absHI}
\centering
\begin{tabular}{lcccc}
\hline\hline
Line & Area  &  V  &  $\Delta$V & Peak  \\
        & \kms\  &  \kms\  &  \kms\ &       \\
\hline
 HI$_1$  & 0.11 $\pm$.01   &-24 $\pm$ 1.8  &  33.  $\pm$1.8  &     0.9967\\
 HI$_2$   & 0.07 $\pm$.01   & -10. $\pm$1.8 &    8   $\pm$1.8   &    0.9913\\
  HI$_3$  & 0.16 $\pm$.01   & -2.6 $\pm$1.8 &  4.5 $\pm$1.8   &  0.9664\\
  HI$_4$  & 0.17 $\pm$.01  &  8.7 $\pm$ 1.8  &  12  $\pm$ 1.8  &   0.9871\\
\hline
OH$_ 1$ &  -0.06$\pm$.01 &   -4 $\pm$2 &   17 $\pm$5  &  0.9966 \\
OH$_2$  & -0.015$\pm$.01 & 14 $\pm$4 &   8  $\pm$6  &  0.9982 \\
\end{tabular}
\begin{list}{}{}
\item -- Area  is the integrated optical depth \\
  -- The $\Delta$V are FWHM; the peaks correspond to the maximum depth of the signal,
  as shown in Fig. \ref{fig:specabs} \\
\end{list}
\end{table}

\subsection{Absorption upper limits}    
\label{sec:absUL}

The wide bandwidth of NOEMA allowed us to search for other possible absorptions
from lines falling in our frequency range. There is an H$_2$CO line at
101.332 GHz; however, the lower level of the transition at 57~K is too high
to yield a significant limit. There is CS(2-1) at 97.980 GHz,
with a lower level at 2.2~K, and the fundamental O$_2$ line at 
 118.750 GHz. These lines were not detected, and we derive  3$\sigma$ upper limits
 of N(CS) $\leq$ 3.3$\times$ 10$^{12}$ cm$^{-2}$ and 
 N(O$_2$) $\leq$ 2.9$\times$ 10$^{17}$ cm$^{-2}$, assuming
 the same excitation temperature of T$_x$ = 15~K. The molecular oxygen limit
 is quite high, because the strength of the transition is about 2 orders
 of magnitude lower than for the CO molecule.

\section{Concluding remarks}    
\label{sec:conc}  

We  reported the discovery of a new CO(1-0) absorption in an intervening
galaxy ($z=0.05$) in front of a background quasar ($z=1.3$).
Millimetric molecular absorptions are still very rare at moderate and 
high redshift: only six systems have been found \citep[e.g.,][]{Combes2008}: 
three are associated-absorbing systems (i.e., from the AGN host itself), one
has been detected recently \citep{Allison2019}, and 
three are intervening absorbers from gravitational lens systems. To date
 no simple millimeter molecular absorber 
has been detected in a normal intervening galaxy without strong lensing, and hence 
more suitable to study variations of fundamental constants. The absorber
in front of Q0248 presented here is thus the first one: the absorbing gas is
in a tidal tail without any lensing potential.

With high spatial resolution, the number of detections for local millimeter 
absorption, in the associated-absorption category, is increasing 
\citep[e.g.,][]{Tremblay2016}, providing extremely useful information, 
for example disentangling inflow from outflow around AGN.
Although intervening \hi\ 21cm absorbers are now more frequent with about a 30\% detection 
rate in case of optically selected sight lines \citep{Gupta2009, Gupta2013, Dutta2017}, 
the OH absorbers in intervening galaxies are still rare, with an incidence
or a number per unit redshift 
of $n_{OH}=dN_{OH}/dz$=0.008, at z$\sim$0.1 \citep{Gupta2018OH}.

In the present case, the absorption is from relatively diffuse gas, belonging
to a tidal tail at about 17~kpc projected distance from the parent G1 galaxy,
the gas-rich spiral from the merging pair. The column density is therefore quite low,
and the depth of the absorption rather small, less than 2\%.  In the near future, the
increased sensitivity of ALMA, NOEMA, and large blind surveys with SKA precursors such
as MALS \citep[e.g.,][]{Gupta2016, Allison2016} will make it possible to discover
such weak absorbing systems, which was impossible before.

\begin{acknowledgements}
  We warmly thank the referee for the constructive comments and suggestions.
  We thank the GMRT and IRAM staff for their support during the observations, and in particular M\'elanie Krips
during the reduction of NOEMA data.  
We acknowledge the use of ARTIP.  ARTIP was developed by researchers and developers at ThoughtWorks 
India Pvt. Limited and IUCAA.  
GMRT is run by the National Centre for Radio Astrophysics of the Tata Institute of Fundamental Research.
The National Radio Astronomy Observatory is a facility of the National Science Foundation operated under cooperative agreement by Associated Universities, Inc.
This work is based on observations carried out under project number W17CE with the
IRAM NOEMA Interferometer. IRAM is supported by INSU/CNRS (France), MPG (Germany), and IGN (Spain).
The Common Astronomy Software Applications (CASA) package was 
developed by an international consortium of scientists based at the National Radio 
Astronomical Observatory (NRAO), the European Southern Observatory (ESO), the National 
Astronomical Observatory of Japan (NAOJ), the Academia Sinica Institute of Astronomy 
and Astrophysics (ASIAA), the CSIRO division for Astronomy and Space Science (CASS), 
and the Netherlands Institute for Radio Astronomy (ASTRON) under the guidance of NRAO. 
\end{acknowledgements}


\bibliographystyle{aa}
\bibliography{g0248-CO}
\end{document}